\documentstyle[12pt]{article}
\begin{document}

\begin{flushright}
UT-Komaba 99-14
\end{flushright}


\begin{center} 
{\Large{\bf Gross-Neveu model with overlap fermions }}\\
\vskip 1.5cm

{\Large  Ikuo Ichinose\footnote{e-mail 
 address: ikuo@hep1.c.u-tokyo.ac.jp}and 
 Keiichi Nagao\footnote{e-mail
 address: nagao@hep1.c.u-tokyo.ac.jp}}  
\vskip 0.5cm
 
Institute of Physics, University of Tokyo, Komaba,
  Tokyo, 153-8902 Japan  
 
\end{center}

\vskip 2cm
\begin{center} 
\begin{bf}
Abstract
\end{bf}
\end{center}
We investigate the chiral properties of overlap lattice fermion by 
using two dimensional Gross--Neveu model coupled with a gauge field.
Chiral properties of this model are similar to those of QCD$_4$, that is, 
the chiral symmetry is spontaneously broken in the presence of 
small but finite fermion mass and also there appears the chiral anomaly
because of the coupling with the gauge field. 
In order to respect L\"uscher's extended chiral symmetry we 
insert overlap Dirac operator even in the interaction terms so that 
the whole action including them are invariant under the extended
chiral transformation
at {\em finite lattice spacing}, though the interaction terms 
become nonlocal.
We calculate mass of the quasi-Nambu--Goldstone boson as 
a function of the bare fermion mass and two parameters in the overlap
formalism, and find that the quasi-Nambu--Goldstone 
boson has desired properties 
as a result of the extended chiral symmetry.
We furthermore examine the PCAC relation and find that it is satisfied 
at {\em finite lattice spacing}. 
Relationship between the anomaly term in the PCAC relation and the U(1)
problem is also discussed.

\newpage
\setcounter{footnote}{0}

Lattice fermion formulation is one of the most important problem in the 
lattice field theory.
Recently a very promising formalism was proposed, the overlap fermion
formalism\cite{NN},
and after that there appeared a lot of works on that.
Importance of the Ginsparg and Wilson (GW) relation\cite{GW} was 
stressed there.

In this paper we shall study or test the overlap fermion by
using the Gross-Neveu model in two dimensions.
This is a solvable model which has 
chiral properties similar to  those of
QCD$_4$, i.e., chiral symmetry is spontaneously
broken with a small but finite bare fermion mass and pion appears
as quasi-Nambu-Goldstone boson\footnote{Overlap fermion with
a finite fermion mass was recently studied in
Refs.\cite{Chandra},\cite{mass},\cite{IN}. 
For study of the domain wall fermion in  
four-Fermi models, see Ref.\cite{VTK}.}.
Furthermore in both models the U(1) problem is solved by the chiral anomaly
which comes from the coupling with the gauge field.

In order to respect L\"uscher's extended chiral symmetry we 
insert the overlap Dirac operator in the interaction
terms so that 
the whole action including them are invariant under the extended
chiral transformation at {\em finite lattice spacing}, 
though the interaction terms 
becomes nonlocal\cite{kiku}.\footnote{Similar 
modification is discussed in 
Refs.\cite{luscher},\cite{chui}.}\footnote{We studied in 
Ref.\cite{IN} the chiral properties of overlap 
fermions by using the same model, where 
the interaction terms are invariant under the {\em ordinary} chiral 
transformation. }\footnote{In the leading order of the $1/N$-expansion,
the nonlocality is expected harmless because of the exponential decay
of the large-distance terms\cite{Locality}.} 
As a result, only the fermion bare mass breaks the extended chiral symmetry
(and the measure of the path-integral of fermions).

The model is given by the following action on the lattice with 
lattice spacing $a$,
\begin{eqnarray}
S&=& N \sum_{pl}[1-\frac{1}{2}(U_P + U_P^\dag)] 
+ a^2\sum_{n,m}\bar{\psi}(m)D(m,n)\psi(n)
-a^2M_B\sum_n\bar{\psi}\psi(n)   \nonumber \\
&& 
-{a^2 \over \sqrt{N}}\sum_n\Big[\phi^i(n)\bar{\psi}(n)\tau^i
\Big(\delta_{nm}-{a\over 2}D(n,m)\Big)\psi(m)  \nonumber  \\
&&  +\phi^i_5(n)\bar{\psi}(n)\tau^i\gamma_5
\Big(\delta_{nm}-{a\over 2}D(n,m)\Big)\psi(m)\Big] \nonumber  \\
&& +{a^2 \over 2g_v}\sum_n\Big[\phi^i(n)\phi^i(n)+
\phi^i_5(n)\phi^i_5(n)\Big],
\label{action1}
\end{eqnarray}
where $U_P$ is the plaquette variable of $U_\mu (n)$, 
U(1) gauge field defined on links,
$\psi^l_{\alpha}\; (\alpha=1,...,N,
l=1,...,L)$ are fermion fields with flavour index $l$,
and the matrix $\tau^i\; (i=0,...,L^2-1)$ acting on the flavour
index is normalized as
\begin{equation}
{\rm{Tr}}(\tau^i\tau^k)=\delta_{ik}
\end{equation}
and 
\begin{equation}
\tau^0={1 \over \sqrt{L}}, \; \; \{\tau^i,\tau^j\}=d^{ijk}\tau^k.
\end{equation}
The covariant derivative of the overlap fermion $D(n,m)$ 
which satisfies the GW relation
\begin{equation}
D\gamma_5+\gamma_5D=aD\gamma_5D.
\label{GWr}
\end{equation}
is defined by
\begin{eqnarray}
D&=&{1\over a}\Big(1+X{1 \over \sqrt{X^{\dagger}X}}\Big),  \nonumber  \\
X_{nm}&=&\gamma_{\mu}C_{\mu}(n,m)+B(n,m),  \nonumber   \\
C_{\mu}(m,n)&=&{1 \over 2a}\Big[\delta_{m+\mu,n}U_{\mu}(m)-
\delta_{m,n+\mu}U^{\dagger}_{\mu}(n)\Big],  \nonumber  \\
B(m,n)&=&-{M_0\over a}+{r\over 2a}\sum_{\mu}\Big[2\delta_{n,m}
-\delta_{m+\mu,n}U_{\mu}(m)-\delta_{m,n+\mu}U^{\dagger}_{\mu}(n)\Big],
\label{covD}
\end{eqnarray} 
where $r$ and $M_0$ are 
dimensionless nonvanishing free parameters of the overlap lattice fermion 
formalism\cite{NN,Neuberger}.
For the vanishing bare fermion mass $M_B=0$, the action (\ref{action1}) 
is invariant under the following extended chiral transformation, 
\begin{eqnarray}
&& \psi(n) \rightarrow \psi(n)+\tau^k\theta^k\gamma_5\Big\{
\delta_{nm}-aD(n,m)\Big\}\psi(m),  \nonumber  \\
&& \bar{\psi}(n) \rightarrow \bar{\psi}(n)+\bar{\psi}(n)
\tau^k\theta^k\gamma_5,  \nonumber  \\
&&\phi^i(n) \rightarrow \phi^i(n)+d^{ikj}\theta^k\phi^j_5(n),  \nonumber  \\
&&\phi^i_5(n) \rightarrow \phi^i_5(n)-d^{ikj}\theta^k\phi^j(n),
\label{extended}
\end{eqnarray}
where $\theta^i$ is an infinitesimal transformation parameter.

From the action (\ref{action1}), $\phi^i$ and $\phi^i_5$
are (nonlocal) composite fields of the fermions,
\begin{equation}
\phi^i={g_v\over \sqrt{N}}\bar{\psi}\tau^i\Big(1-{a \over 2}D\Big)
\psi, \;\; 
\phi^i_5={g_v\over \sqrt{N}}\bar{\psi}\tau^i\gamma_5\Big(1-{a \over 2}D\Big)
\psi.
\label{eq.motion}
\end{equation}
As in the continuum model, we expect that the field
$\phi^0$ acquires a nonvanishing vacuum expectation value (VEV)
\begin{equation}
\langle\phi^0\rangle=\sqrt{NL}M_s,
\label{VEV}
\end{equation}
and we define subtracted fields,
\begin{eqnarray}
&&\varphi^0=\phi^0-\sqrt{NL}M_s,  \nonumber  \\
&&\varphi^i=\phi^i\; \; (i\neq 0),  \;\; \varphi^i_5=\phi^i_5.
\end{eqnarray}
In Ref.\cite{Chandra}, it is argued that the nonlocal
composite field $\phi^0$ works as an order parameter
for the extended chiral symmetry.
Then the action is rewritten as,
\begin{eqnarray}
S&=& N \sum_{pl}[1-\frac{1}{2}(U_P + U_P^\dag)]
+ a^2\sum_{n,m}\bar{\psi}(m)D'_M(m,n)\psi(n)   \nonumber \\
&& 
-{a^2 \over \sqrt{N}}\sum_n\Big[\varphi^i(n)\bar{\psi}(n)\tau^i
\Big(\delta_{nm}-{a\over 2}D(n,m)\Big)\psi(m)  \nonumber  \\
&& +\varphi^i_5(n)\bar{\psi}(n)\tau^i\gamma_5
\Big(\delta_{nm}-{a\over 2}D(n,m)\Big)\psi(m)\Big] \nonumber  \\
&& +{a^2 \over 2g_v}\sum_n\Big[\varphi^i(n)\varphi^i(n)+
2\sqrt{NL}M_s\varphi^{0}(n)+\varphi^i_5(n)\varphi^i_5(n)\Big],
\label{action2}
\end{eqnarray}
where 
\begin{eqnarray}
&& D'_M(n,m)=D'(n,m)-M\delta_{nm},  \nonumber  \\
&& D'(n,m)=cD(n,m),   \nonumber   \\
&& M=M_B+M_s,  \;\;\; c=1+{M_sa\over 2}.
\end{eqnarray}

From the extended chiral symmetry (\ref{extended})
and its spontaneous breaking (\ref{VEV}), we can expect that
quasi-Nambu--Goldstone bosons appear. 
They are nothing but $\phi^i_5$.
However there is a subtle problem for the Goldstone theorem because of the 
nonlocality of the action.
But as we explaned above, this nonolcality of the action is
expected to be harmless in the leading order of the $1/N$-expansion.
We shall explicitly examine this point.

The VEV $M_s$ is determined by the tadpole cancellation condition of
 $\phi^0$. In order to perform an explicit calculation of the
 $1/N$-expansion, it is useful to employ the momentum representation and
 also we introduce the gauge potential $\lambda_\mu(n)$ in the usual
 way, i.e., $U(n,\mu)=\exp ({ia\over \sqrt{N}}\lambda_{\mu}(n))$.
By using weak-coupling expansion by Kikukawa and Yamada\cite{weakcoupling},
\begin{eqnarray}
D_{nm}&=&\int_p\int_qe^{-ia(qn-pm)}D(p,q),    \\
D(p,q)&=&D_0(p)(2\pi)^2\delta(p-q)+{1 \over a}V(p,q),
\label{overlapF}
\end{eqnarray} 
where $\int_p=\int^{\pi/a}_{-\pi/a}{d^2p\over (2\pi)^2}$
and 
\begin{eqnarray}
D_0(p)&=& {b(p)+\omega(p) \over a\omega(p)}
+{\gamma_\mu i \sin ap_\mu \over a^2\omega(p)},  \label{D0}  \\
V(p,q) &=& 
 \Bigl\{ 
\frac{1}{\omega(p)+\omega(q)} 
\Bigr\}
\Bigl[
X_1(p,q)
-
\frac{X_0(p)}{\omega(p)} X^\dagger_1(p,q) 
\frac{X_0(q)}{\omega(q)}
\Bigr] +...
\end{eqnarray}
\begin{eqnarray}
X_0(p)&=& 
\frac{i}{a} \gamma_\mu \sin a p_\mu 
+ \frac{r}{a} \sum_\mu \left(1-\cos a p_\mu \right) 
-\frac{1}{a} M_0 ,\\
&& \nonumber\\
X_1(q,p)&=& \int_k  (2\pi)^4 \delta(q-p-k)
\, \frac{1}{\sqrt N} \lambda_\mu(k) \, V_{1 \mu}\left(p+\frac{k}{2}\right) , 
\end{eqnarray}
\begin{eqnarray}
a\omega(p)&=& \sqrt{\sin^2(ap_\mu)
+\Big(r\sum_\mu(1-\cos(ap_\mu))-M_0\Big)^2},  \nonumber   \\
ab(p) &=& r\sum_\mu(1-\cos (ap_\mu))-M_0.
\end{eqnarray}
The vertex function is explicitly given as 
\begin{eqnarray}
V_{1 \mu}\left(p+\frac{k}{2}\right) 
&=& i \gamma_\mu \cos a \left(p_\mu+\frac{k_\mu}{2}\right) 
+ r \sin a \left(p_\mu+\frac{k_\mu}{2}\right) \nonumber \\
&=& \frac{\partial}{\partial p_\mu} X_0 \left(p+\frac{k}{2}\right).
\end{eqnarray}
 
From (\ref{D0}), the tree level propagator is obtained as
\begin{eqnarray}
D'^{-1}_{M(0)}&=& {a\{cb(p)+(c-Ma)\omega(p)\} -ic\gamma_{\mu}
\sin (ap_\mu) \over \omega(p)\{c^2+(c-Ma)^2\}+2cb(p)(c-Ma)}  \nonumber  \\
&\equiv& {A^c_{\mu}(p)\gamma_\mu+B^c(p) \over J^c(p)},
\label{propa}
\end{eqnarray}
where
\begin{eqnarray}
a\omega(p)&=& \sqrt{\sin^2(ap_\mu)
+\Big(r\sum_\mu(1-\cos(ap_\mu))-M_0\Big)^2},  \nonumber   \\
ab(p) &=& r\sum_\mu(1-\cos (ap_\mu))-M_0.
\end{eqnarray}
From (\ref{action2}) and (\ref{propa}), the VEV is determined as 
follows by the tadpole cancellation condition, 
\begin{eqnarray}
{M_s  \over g_v}&=&-\int_k\mbox{Tr}\Big[D_{M(0)}^{'-1}(1-{a\over 2}
D_0(k))\Big]  \nonumber   \\
&=&\int_k{Ma^2(\omega(k)-b(k)) \over J^c(k)}.
\label{Ms}
\end{eqnarray}

Effective action of $\varphi^i$, $\varphi^i_5$ and the gauge field
$\lambda_\mu(n)$ is obtained by integrating over the 
fermions,
\begin{equation}
  e^{-S_{eff}}=\int [D\bar{\psi}D\psi] e^{-S}.
\end{equation}
For the quasi-Nambu-Goldstone boson $\varphi_5$,  
\begin{equation}
S^{(2)}_{eff}[\varphi_5]=\int_p{1 \over 2}
\varphi^i_5(-p)\Gamma^5_{ij}(p)\varphi^j_5(p)
\label{Svphi}
\end{equation}
where
\begin{eqnarray}
\Gamma^5_{ij}(p)&=&\delta_{ij}\Big\{{1\over g_v}+\int_k
\mbox{Tr}\Big[\gamma_5\Big\{1-{a\over 2}D_0(k-p)\Big\}
\langle \psi(k-p)\bar{\psi}(k-p)\rangle \gamma_5  \nonumber  \\
&& \;\; \times\Big\{1-{a\over2}D_0(k)\Big\}\langle \psi(k)\bar{\psi}(k)\rangle
  \Big]\Big\}  \nonumber   \\
&=&\delta_{ij}[\epsilon +2M_0^2\hat{A}(p;M)],
\label{Gamma5}
\end{eqnarray}  
$\hat{A}(p=0)=0$ and 
\begin{eqnarray}
\epsilon&=&{1\over g_v}+\int_k
\mbox{Tr}\Big[\gamma_5\Big\{1-{a\over 2}D_0(k)\Big\}
\langle \psi(k)\bar{\psi}(k)\rangle \gamma_5  \nonumber  \\
&& \;\; \times\Big\{1-{a\over2}D_0(k)\Big\}\langle \psi(k)\bar{\psi}(k)\rangle
  \Big]  \nonumber   \\
&=&{M_Ba^2\over M_s}\int_k {\omega(k)-b(k) \over J^c(k)}\nonumber \\
&=& -{M_B \over M M_s}\int_k\mbox{Tr}\Big[D_{M(0)}^{'-1}(1-{a\over 2}
D_0(k))\Big].  
\label{epsilon}
\end{eqnarray}
$M_s$ in Eq.(\ref{Ms}) represents the order parameter of 
the extended chiral symmetry 
at {\em finite lattice spacing} as suggested by 
Chandrasekharan\cite{Chandra}.

The momentum dependent term in (\ref{Gamma5}) has a rather
complicated form at finite lattice spacing but in the continuum
limit it coincides with the previous result,
\begin{eqnarray}
\hat{A}(p;M)&\rightarrow& {p^2 \over 4\pi\sqrt{p^2(p^2+\mu^2)}}
\ln {p^2+2\mu^2+\sqrt{(p^2+2\mu^2)^2-4\mu^4} \over 
p^2+2\mu^2-\sqrt{(p^2+2\mu^2)^2-4\mu^4}}   \nonumber  \\
&\rightarrow& {p^2 \over 4\pi\mu^2}+O((p^2)^2).
\label{Ak}
\end{eqnarray}
where $\mu=M_0M$. 
In a similar way, the mixing term of the gauge field and the 
flavour-singlet ``pion" $\varphi^0_5$ is given in the continuum limit
as
\begin{equation}
S^{(2)}_{eff}[\lambda_\mu,\varphi^0_5] =-2\sqrt{L}
M^2_0M\int_p\sum \lambda_\mu(-p)
{\epsilon_{\mu\nu}p_\nu \over p^2} \hat{A}(p;M)\varphi^0_5(p).
\label{mixing}
\end{equation}
This mixing term is related with the chiral anomaly and it coincides with
the standard form only in the continuum limit.
This indicates that the chiral anomaly which appears from the measure
of the fermion path integral coincides with its standard form only
in the continuum limt.\footnote{However there is some argument
claiming that the chiral anomaly of the standard form appears even
at the finite lattice spacing\cite{Lanomaly} (see also later discussion
on the PCAC relation).}

We shall calculate the masses of the fields $\varphi^i_5$
by integrating out the gauge field,
\begin{eqnarray}
&&\int [DU]\exp\Big[-S_G[U] - S_{eff}[\lambda_\mu,\varphi_5] 
-S_{eff}[\varphi_5]\Big] \nonumber \\ \nonumber
&=&\int \prod d\lambda_1(p)\exp\Big[-\int_p \frac{1}{2}p_2^2 \lambda_1(p)\lambda_1(-p)+2\sqrt{L}M^2_0M\int_p\sum \lambda_\mu(-p)
{\epsilon_{\mu\nu}p_\nu \over p^2} \hat{A}(p;M)\varphi^0_5(p)\\ \nonumber
&& -\int_p{1 \over 2}
\varphi^i_5(-p)\Gamma^5_{ij}(p)\varphi^j_5(p)\Big]  \\ \nonumber
&=& \int \prod d\lambda_1(p)\exp\Big[-\int_p \frac{1}{2}p_2^2 \{(\lambda_1(p)
-\frac{\sqrt L}{2\pi M p_2}\varphi_5^0(p))(\lambda_1(-p)+
\frac{\sqrt{L}}{2\pi M p_2}\varphi_5^0(-p)) \\ \nonumber
&&+ \frac{L}{(2\pi M)^2 p_2^2}\varphi_5^0(p)\varphi_5^0(-p)
+ \frac{1}{p_2^2}\varphi_5^0(-p)\Gamma^5(p)\varphi_5^0(p)\}\Big]\\ \nonumber
&&\times \exp\Big[-\int_p {1 \over 2}
\varphi^i_5(-p)\Gamma^5_{ij}(p)\varphi^j_5(p)\Big]\\ 
&\propto& \exp\Big[-\frac{1}{2}\int_p \varphi_5^i(p)\times 
\frac{2\pi p^2+(2\pi M)^2 \epsilon+L\delta^{i0}}{(2\pi M)^2}
 \times \varphi_5^i(-p)\Big]
\end{eqnarray}
where we took axial gauge fixing($\lambda_2 =0$). 
Therefore we get the mass of eta $\varphi_5^0$ and pions 
$\varphi_5^i(i\neq 0)$ as follows
\begin{equation}
m_\eta ^2 =m_\pi ^2 + \frac{L}{2\pi} ,
\end{equation}
\begin{equation} 
m_\pi ^2 = 2\pi M^2 \epsilon .
\end{equation}
It is obvious that the mass of the quasi-NG bosons, the
flavour-non-singlet pions $\varphi_5^i(i\neq 0)$,
 is proportional to the bare fermion mass $M_B$
and vanishes as $M_B\rightarrow 0$\cite{Hasenfratz}.
This is the expected result.
In the leading order of the $1/N$-expansion, the gauge variables in the
derivative $D(m,n)$ are set as $U_\mu (n)=1$.
Then from Ref.\cite{Locality}, $D(n,m)$ decays exponentially as 
$|n-m|^2$ tends to large.\footnote{The Goldstone theorem assumes
the locality of the current.
For the overlap fermion, see Ref.\cite{KY}.}
On the other hand the flavour-singlet meson $\varphi_5^0$ has
the finite mass $\sqrt{L\over 2\pi}$ even in the limit $M_B\rightarrow 0$.
This means that the U(1) problem is solved by the existence of the
mixing term (\ref{mixing}).
Relationship between the mixing term and the chiral anomaly
will be discussed after study of the PCAC relation.

Let us turn to the Ward-Takahashi (WT) identity for the axial-vector
current; the PCAC relation.
We perform the following change of variables in the functional integral
of the partition function,
\begin{eqnarray}
&& \psi(n) \rightarrow \psi(n)+\tau^k\theta^k(n)\gamma_5\Big\{
\delta_{nm}-aD(n,m)\Big\}\psi(m),  \nonumber  \\
&& \bar{\psi}(n) \rightarrow \bar{\psi}(n)\left\{1+\tau^k\theta^k(n)\gamma_5
\right\},  \nonumber  \\
&&\phi^i(n) \rightarrow \phi^i(n)+d^{ikj}\theta^k(n)\phi^j_5(n),  \nonumber  \\
&&\phi^i_5(n) \rightarrow \phi^i_5(n)-d^{ikj}\theta^k(n)\phi^j(n),
\label{extended2}
\end{eqnarray}
and we obtain the following WT identity,
\begin{equation}
\langle \nabla_\mu J^k_{5,\mu}(n)-{2\sqrt{N} \over g_v}
M_B\varphi^k_5(n)-\delta^{k0}N\sqrt{L}a\mbox{Tr}\Big[\gamma_5D(n,n)
\Big]\rangle=0,
\label{WT1}
\end{equation}
where the last term comes from the measure of the fermion path integral and 
$\nabla_\mu$ is the difference operator on the lattice.
The axial vector current $J^k_{5,\mu}(n)$ is given by
\begin{equation} 
J^k_{5,\mu}(n)=\sum_{lm}\bar{\psi}(l)\Big[c\tau^k+
{a \over 2\sqrt{N}}(\varphi^i(n)+\varphi^i_5(n)\gamma_5
+\sqrt{NL}M_s\delta_{i0})\tau^i\tau^k\Big]
K^5_{n\mu}(l,m)\psi_m.
\label{AVC}
\end{equation}
The terms proportional to the boson fields and $M_s$ in
Eq.(\ref{AVC}) come from the interaction terms
of the fermions and bosons in the action (\ref{action2}) and  
explicit form of $K^5_{n\mu}(l,m)$ was obtained by Kikukawa and 
Yamada\cite{KY} as follows,
\begin{eqnarray}
K^5_{n\mu}(l,m)&=&\left\{ K_{n\mu}\frac{H}{\sqrt{H^2}}\right\}(l,m),\\ 
H&=&-\gamma_5 X, 
\end{eqnarray}
\begin{equation}
a K_{n\mu}(l,m)
= \gamma_5 \left\{
\int_{-\infty}^{\infty} \frac{dt}{\pi}
\frac{1}{ (t^2+H^2)}
\left( t^2 W_{n\mu} - H W_{n\mu} H \right)
\frac{1}{ (t^2+H^2) } 
\right\}_{lm}, 
\end{equation}
\begin{equation}
W_{n\mu}(l,m)
= \gamma_5 \left\{ 
\frac{1}{2} \left(\gamma_\mu-1\right)
                 \delta_{nl}\delta_{n+\hat\mu, m} \, U_{n\mu}
+ \frac{1}{2} \left(\gamma_\mu+1\right)
 \delta_{l,n+\hat\mu}\delta_{nm} \, U_{n+\hat\mu,\mu}^\dagger 
\right\}. \nonumber\\
\end{equation}
From (\ref{Ms}) and (\ref{epsilon}), we have
\begin{equation}
{1\over g_v}={M\over M_B}\; \epsilon.
\label{gvepsilon}
\end{equation}
Then the WT identity (\ref{WT1}) gives 
the following PCAC relation in the continuum limit
\begin{eqnarray}
\partial_\mu J^k_{5,\mu}&=&i\delta^{k0}{\sqrt{NL} \over \pi}
\sum_{\mu\nu}\epsilon_{\mu\nu}\partial_\nu\lambda_\mu+2M\epsilon\sqrt{N}
\varphi^k_5  \nonumber   \\
&=&i\delta^{k0}{\sqrt{NL} \over \pi}
\sum_{\mu\nu}\epsilon_{\mu\nu}\partial_\nu\lambda_\mu+\sqrt{{2N\over \pi}}
m^2_\pi\times {\varphi^k_5\over \sqrt{2\pi M^2}}.
\label{WT2}
\end{eqnarray}
In Eq.(\ref{WT2}), the anomaly term comes from the last term in
Eq.(\ref{WT1}).
Then it is obvious that the desired PCAC relation (besides the anomaly 
term which we obtained in the continuum limit
\footnote{It is possible to obtain the anomaly term at finite lattice 
spacing \cite{Lanomaly}, but in this paper we calculated this term 
in the continuum limit.})
is obtained rather straightforwardly from (\ref{WT1}) and 
(\ref{gvepsilon}) at {\em finite lattice spacing}.

Let us briefly discuss relationship between the U(1) problem and 
the chiral anomaly.
As we saw, the U(1) problem is solved by the mixing term in the 
effective action (\ref{mixing}).
It is obvious that in $S_{eff}$ 
there exist two terms which (explicitly) break
the extended chiral symmetry, i.e., the mass term of $\varphi^i_5$
and the mixing term.
According to the nonvanishing VEV 
$\langle\phi_0\rangle=\sqrt{NL} M_s$ (Eq.(\ref{VEV})),
$\varphi^i_5$ transform as 
\begin{equation}
\varphi^i_5 \rightarrow \varphi^i_5-{1\over \sqrt{L}}\theta^i
\langle\phi_0\rangle +\cdots
\end{equation}
under the extended chiral transformation.
Then it is obvious that the above two terms in the effective action
generate terms in the WT identity in the level of the effective action
which correspond to the two terms
on the RHS of Eq.(\ref{WT2})\cite{KO,Kawa}.

In this paper we studied the overlap fermion by using the two dimensional 
gauged Gross-Neveu model in the large $N$ limit 
and showed that the pion mass is proportional to the bare fermion
mass and that the PCAC relation is satisfied in the desired form
even at finite lattice spacing.
Relationship between the anomaly term in the PCAC relation and the U(1)
problem is also discussed.
We expect that the extended chiral symmetry works properly
as a genuine continuous symmetry and that the nonlocality of the 
transformation and the action does not generate any serious issues
on the physical results at least in the perturbative expansion
of the gauge field.
In order to examine this expectation, studies of the higher-order
terms of $1/N$ is useful.
In this paper we also identified the order parameter for
the extended chiral symmetry and the pion field.
This is useful for studies of QCD with overlap fermions which is 
invariant under the extended chiral symmetry\cite{IN-QCD}.

\bigskip
{\bf Acknowledgments}  \\
KN would like to thank Prof.Y.Kikukawa for 
useful discussions about the interaction terms which are invariant 
under the extended chiral symmetry, and 
also would like to thank Prof.T.Yoneya for helpful comments.
Furthermore II and KN would like to thank Prof.K.Kawarabayashi for 
useful discussions about U(1) problem.

\end{document}